\documentclass[aps,pre,reprint]{revtex4-1}
\usepackage{times}		
\usepackage{booktabs}
\usepackage{placeins}

\usepackage{amssymb,amstext,amsmath, epsf}
\usepackage{graphicx,graphics,epsfig,wrapfig}
\usepackage{listings}
\usepackage{float}
\usepackage{esint}

\usepackage{color}
\usepackage[english]{babel}
\usepackage[T1]{fontenc}
\usepackage[utf8]{inputenc}

\def\apj{{ApJ}}
\def\apjl{{ApJL}}

\def\nat{{Nature Physics}}
\def\prl{{PRL}}

\def\pre{{PRE}}

\def\aap{{Astronomy and Astrophysics}}
\def\physscr{{Physica Scripta}}

\begin{document}

\title{(Quasi-)collisional Magneto-optic Effects in Collisionless Plasmas with sub-Larmor-scale Electromagnetic Fluctuations}

\author{Brett D. Keenan}
\email{bdkeenan@ku.edu}
\author{Alexander L. Ford}
\affiliation{Department of Physics and Astronomy, University of Kansas, Lawrence, KS 66045}
\author{Mikhail V. Medvedev}
\affiliation{Department of Physics and Astronomy, University of Kansas, Lawrence, KS 66045}


\begin{abstract}
High-amplitude, chaotic/turbulent electromagnetic fluctuations are ubiquitous in high-energy-density laboratory and astrophysical plasmas, where they can be excited by various kinetic-streaming and/or anisotropy-driven instabilities, such as the Weibel instability. These fields typically exist on ``sub-Larmor scales'' --- scales smaller than the electron Larmor radius. Electrons moving through such magnetic fields undergo small-angle stochastic deflections of their pitch-angles, thus establishing diffusive transport on long time-scales. We show that this behavior, under certain conditions, is equivalent to Coulomb collisions in collisional plasmas. The magnetic pitch-angle diffusion coefficient, which acts as an effective ``collision'' frequency, may be substantial in these, otherwise, collisionless environments. We show that this effect, colloquially referred to as the plasma ``quasi-collisionality'', may radically alter the expected radiative transport properties of candidate plasmas. We argue that the modified magneto-optic effects in these plasmas provide an attractive, novel radiative diagnostic tool for the exploration and characterization of small-scale magnetic turbulence, as well as affect inertial confinement fusion and other laser-plasma experiments.
\end{abstract}

\maketitle

\section{Introduction}
\label{s:intro}

Strong electromagnetic turbulence is a common phenomenon in high-energy-density (HED) environments. In the laboratory settings, manipulating and understanding electromagnetic turbulence is essential to fusion energy science and the inertial confinement fusion (ICF) \citep{ren04, tatarakis03}. In addition, electromagnetic turbulence is critical to numerous astrophysical systems such as gamma-ray bursts and supernova shocks \citep{medvedev09b, medvedev06, reynolds12, kamble+14}, as well as in laboratory astrophysics laser-plasma experiments\citep{huntington15,park15}.

Despite much variation in the origin of the electromagnetic turbulence (e.g., the Weibel or filamentation instabilities), most of these plasmas have one thing in common: their configuration is such that binary Coulomb collisions are negligible; i.e., the plasmas are ``collisionless''. Nonetheless, some of these environments, such as plasmas at ``collisionless'' shocks, display phenomena that resemble conventional collisional interactions. Hereafter, we colloquially refer to these phenomena as ``quasi-collisional''. 

In this work, we will show that sub-Larmor-scale (``small-scale'') magnetic turbulence induces particle dynamics reminiscent of binary Coulomb interactions. In fact, as we will demonstrate, the random small-angle deflections of electrons by small-scale magnetic fields leads to an effective collisionality with the effective ``collision'' frequency being equal to the (small-angle) pitch-angle diffusion coefficient.

The rest of the paper is organized as follows. Section \ref{s:pitching_it} briefly reviews the analytic theory of pitch-angle diffusion in small-scale random magnetic fields. Next, we argue that the small-angle deflections, characteristic of these fields, are analogous to the small deflections induced by Coulomb collisions. We then show that the pitch-angle diffusion coefficient, itself, acts as an effective collision frequency. In Section \ref{s:coll_eqs}, we explore the implications for electromagnetic wave propagation in magnetized plasmas with high ``effective collisionality'' or ``quasi-collisionality''. Finally, Section \ref{s:concl} is the conclusions. Unless otherwise specified, we use cgs units throughout the paper.

\section{Small-scale Magnetic Turbulence And Effective Collisionality}
\label{s:pitching_it}

Magnetic fluctuations are known to occur on various spatial scales. We define the fluctuation scale as sub-Larmor if the electron's (fluctuation) Larmor radius, $r_L \equiv \gamma_e\beta m_e c^2/e \langle\delta B\rangle$ is greater than, or comparable to, the spatial correlation length, $\lambda_B$, i.e., $r_L\gtrsim \lambda_B$. Here $\beta=v/c$ is the dimensionless particle velocity, $\langle\delta B\rangle$ is the rms value of the fluctuating field, $m_e$ is the electron mass, $c$ is the speed of light, $e$ is the electric charge, and $\gamma_e$ is the electron's Lorentz factor.

Formally, the correlation length is defined over all spatial scales of the magnetic field. Nonetheless, any realization of electromagnetic turbulence may be envisioned as the superposition of ``small-scale'' and ``large-scale'' (i.e., the ``sub-'' and ``super-Larmor-scale'') components. Thus, we may roughly define two characteristic spatial scales for the general case, where $\lambda_B^{ssc}$ and $\lambda_B^{lsc}$ are the sub-Larmor-scale and super-Larmor-scale correlation lengths, respectively.

Ignoring the mean magnetic field, there are a number of different regimes that may be enumerated, depending upon the relative significance of the magnetic field at each scale. Firstly, if the correlation length is infinite, then the electrons will follow helical orbits about the axis of a perfectly homogeneous magnetic field. Next, if the magnetic field is ``large-scale'' --- i.e., possessing fluctuations on a finite, though super-Larmor, spatial scale --- then the electron's guiding center will drift, due to slight inhomogeneity in the magnetic field. 

Thirdly, an electron moving through purely sub-Larmor-scale magnetic turbulence will not complete a Larmor orbit, because the magnetic field varies on a scale shorter than the Larmor curvature radius. With $r_L/\lambda_B \gg 1$, this trajectory is a  nearly straight line, with small, random (diffusive) deflections perpendicular to the direction of motion. Finally, when a range of spatial scales exists, the chaotic trajectory will be a combination of large-scale gyro-motions (though not necessarily complete gyro-orbits) with small-scale diffusive deflections. 

We argue that it is these small-scale deflections that induce a {\em quasi-collisionality} with the pitch-angle diffusion coefficient acting as an effective collision frequency.

\subsection{Pitch-angle Diffusion in sub-Larmor-scale Magnetic Turbulence}

Consider an electron moving through a random magnetic field with the mean value $\langle {\bf B} \rangle$. The total magnetic field can be written as:
\begin{equation}
{\bf B}({\bf x}, t) = {\bf B}_0 + \delta{\bf B}({\bf x}, t),
\label{mag_def}
\end{equation} 
where ${\bf B}_0 \equiv \langle {\bf B} \rangle$ and $\delta{\bf B}({\bf x}, t)$ is the mean-free, ``fluctuation'' field, i.e.\ $\langle \delta{\bf B} \rangle = 0$, but $\langle\delta B\rangle\equiv\langle \delta{B}^2 \rangle^{1/2} \neq 0$. 

The pitch-angle diffusion coefficient, due to deflections in purely small-scale magnetic turbulence, is a known function of statistical parameters. It may be obtained by considering that the electron's pitch-angle experiences only a slight deflection, $\alpha_\lambda$, over a single magnetic correlation length. Consequently, the ratio of the change in the electron's transverse momentum, $ \Delta p_\perp$, to its initial momentum, $p$, is $\alpha_\lambda \approx \Delta p_\perp/p \sim e(\delta B/c)\lambda_B/\gamma_e m_e v$, since $\Delta p_\perp \sim F_L\tau_\lambda$ -- where ${\bf F}_L=(e/c)\,{\bf v\times \delta B}$ is the transverse Lorentz force and $\tau_\lambda \sim \lambda_B/v$ is the time to transit $\lambda_B$. The subsequent deflection will be in a random direction, because the field is uncorrelated over the scales greater than $\lambda_B$. As for any diffusive process, the mean squared pitch-angle grows linearly with time. Thus, the diffusion coefficient appears as \citep{keenan13, keenan15}:
\begin{equation}
D_{\alpha\alpha} \equiv \frac{\langle \alpha^2 \rangle}{t} \sim \left(\frac{e^2}{m_e^2c^3}\right)\frac{\lambda_B^{ssc}({\bf x}, t)}{\gamma_e^2\langle \beta^2 \rangle^{1/2}}{\langle \delta B_\perp^2 \rangle},
\label{Daa}
\end{equation}
where $\alpha$ is the electron deflection angle (pitch-angle) with respect to the electron's initial direction of motion, $\delta{{\bf B}_\perp}$ is the component of the fluctuation field perpendicular to the electron's velocity, and $\langle \beta^2 \rangle^{1/2}$ is an appropriate ensemble-average over the electron velocities.

In general, any anisotropy in the fluctuation field will induce a path-dependent correlation length. Thus, the diffusion coefficient along an axis of anisotropy (which is usually along the direction of the mean magnetic field, ${\bf B}_0$) may differ from that across the transverse plane. 

For simplicity, unless otherwise specified, we will assume that the magnetic turbulence is statistically homogeneous and isotropic. With this assumption, the pitch-angle diffusion coefficient will be the same along all directions, and thus we may arbitrary define the axis of the deflection angle, $\alpha$. Without loss of generality, we may then define $\alpha$ as the conventional pitch-angle, i.e., the angle of the velocity vector with respect to the mean (ambient) magnetic field, ${\bf B}_0$. 

Furthermore, we will assume that all relevant time-scales (e.g., the time to transit $\lambda_B$) are much smaller than the magnetic field variability time-scale --- thus, we may treat the magnetic turbulence statically, thereby ignoring any time-dependence in the correlation length, and therefore in $D_{\alpha\alpha}$.

\subsection{The Lorentz Collision Model of Electron-ion Collisions}

In the typical treatment, Coulomb collisions are considered in the small deflection angle regime. In this approximation, a ``test'' electron will undergo a slight (transverse) deflection as it passes by an ion. Additionally, electron-electron collisions are neglected.

Many scatterings will occur, as the binary collisions continue with subsequent ions. These scatterings are effectively stochastic, if the background of (stationary) ions is randomly distributed. Since the collisions with the fixed ion background are elastic, the total electron energy is conserved.

Nevertheless, the small deflections accumulate, leading to a gradual change in the electron's transverse momentum, $\Delta{p_{\perp}}$. An electron is deflected by one radian, i.e. $\Delta{p_{\perp}}/p \sim 1$, in a single collision time, $\tau_c$. The inverse of the collision time, $\nu_{ei} \equiv \tau_c^{-1}$, is defined as the electron-ion collision frequency. 

Given a Maxwellian distribution of electrons, the electron-ion collision frequency assumes the simple form \citep{kruer88}:
\begin{equation}
\nu_{ei} \simeq 3\times10^{-6} \ \text{ln}(\Lambda) \frac{n_e{Z_i}}{\theta_\text{eV}^{3/2}} \ [s^{-1}],
\label{nu_ei_def}
\end{equation}
where $n_e$ is the electron number density in $cm^{-3}$, $\theta_\text{eV}$ is the electron temperature in units of electron-volts, $Z_i$ is the atomic ionization number,  and $\text{ln}(\Lambda)$ is the Coulomb logarithm.

Here, we employ the Spitzer result for the Coulomb logarithm \citep{spitzer56}:
\begin{equation}
\text{ln}(\Lambda) \approx 25.28 + \ln\!\left[\frac{\theta_\text{eV}}{\sqrt{n_e}}\right],
\label{col_log}
\end{equation}
which is valid for temperatures above $4\times10^5 \ K \approx 34 \ eV$.

Next, we will argue that the pitch-angle diffusion coefficient of Eq.\ (\ref{Daa}) acts as an effective collision frequency in plasmas with sub-Larmor-scale magnetic fluctuations.

\subsection{Pitch-angle Diffusion as Effective Collisionality}

The small-angle magnetic deflections are analogous to electron-ion collisional deflections in a number of ways, namely they both (i) conserve particle's energy and (ii) induce deflections transverse to the initial electron's velocity.

Where the two effects differ, however, is in the nature of the stochasticity. In an idealized scenario, an electron in a collisional plasma is continuously deflected by ions along its trajectory. In contrast, an electron moving through small-scale magnetic turbulence is deflected on a characteristic spatial scale of finite length: the correlation length. Thus, the two descriptions are only equivalent on a coarse-graining. Indeed, the electron motion in small-scale magnetic turbulence resembles electron-ion collisions only on spatial scales much greater than the magnetic correlation length. 

Thus, we must require that:
\begin{equation}
L \gg \lambda^{ssc}_B,
\label{size_def}
\end{equation}
where $L$ is the characteristic length scale of the system. 

Next, we may infer this effective collision frequency directly from Eq.\ (\ref{Daa}). The pitch-angle deflections are assumed to be small, hence $\alpha \sim \Delta{p}_\perp/p$. Thus, at $\tau_c$, the following condition must hold:
\begin{equation}
D_{\alpha\alpha}\tau_c \sim 1.
\label{coll_def_mag}
\end{equation}
Therefore, $D_{\alpha\alpha}$ must be the effective ``collision'' frequency.

In general, electron-ion collisions in plasmas are often important too, hence we include them in our study. Consequently, we define the total (effective) collision frequency as: 
\begin{equation}
\nu_\textrm{eff} \equiv \nu_{ei} + D_{\alpha\alpha}.
\label{coll_eff}
\end{equation}

\subsection{A Phenomenological Interpretation}

Electrons undergoing collisions with an ion background will emit Bremsstrahlung radiation. The emission coefficient, $j_\omega$ --- the radiant power per unit frequency per unit volume per unit solid-angle --- is directly proportional to the collision frequency. For a Maxwellian (thermal) distribution of electrons in a weakly ionized plasma, the emission coefficient is \citep{bekefi66}
\begin{equation}
j^\textrm{Brems}_\omega = \Re[n]\left(\frac{\omega_{pe}^2k_BT_e}{8\pi^3c^3}\right)\nu_{ei},
\label{comb_emiss}
\end{equation}
where $\Re[n]$ is the real part of the plasma's index of refraction, $\omega_{pe}= \sqrt{4\pi n_e{e^2}/m_e}$ is the electron plasma frequency, $k_B$ is the Boltzmann constant, and $\nu_{ei}$ is an electron-ion collision frequency. Now, taking into account quasi-collisions as in Eq. \ (\ref{coll_eff}), by substituting $\nu_{ei}\to\nu_\textrm{eff}$ in Eq.\ (\ref{comb_emiss}), the latter introduces a phenomenological definition for the effective collision frequency.

In a similar fashion, electrons undergoing pitch-angle diffusion in small-scale magnetic turbulence emit small-angle jitter radiation 
\citep{medvedev00, medvedev06, medvedev11, RK10, TT11, keenan13, keenan15}. The total (dispersion free) jitter power emitted by a single electron is given by the Larmor formula \citep{keenan15}, i.e.,
\begin{equation}
P_\textrm{tot}^\textrm{jitter}  = {c}\beta r_e^2\gamma_e^2 \langle \delta B_\perp^2 \rangle
\label{jitter_power}
\end{equation} 
where $r_e = e^2/m_e c^2$ is the classical electron radius. The small-angle jitter radiation spectrum has a characteristic frequency known as the jitter frequency,
\begin{equation}
\omega_{j} = \gamma_e^2k_\textrm{mag}\beta{c},
\label{jitter_freq}
\end{equation} 
where $k_\text{mag}$ is the dominant wave number of the (small-scale) turbulent fluctuations. Next, we may write the spectral power for a single electron as:
\begin{equation}
P_\textrm{jitter}(\omega) \equiv \frac{dP}{d\omega} \sim \frac{P_\textrm{tot}^\textrm{jitter}}{\omega_j}.
\label{spec_pow_def}
\end{equation} 
Substitution of Eq.\ (\ref{jitter_freq}) into Eq.\ (\ref{jitter_power}), results in the expression:
\begin{equation}
P_\textrm{jitter}(\omega) \sim \lambda^{ssc}_B \beta \left(\frac{e^4}{m_e^2c^4}\right) \langle \delta B_\perp^2 \rangle,
\label{spec_pow}
\end{equation} 
where the relation, $k_\textrm{mag}^{-1} \sim \lambda^{ssc}_B$, has been employed \citep{keenan15}. Comparing this result to Eq.\ (\ref{Daa}), we find that the power spectrum is directly proportional to the pitch-angle diffusion coefficient:
\begin{equation}
P_\textrm{jitter}(\omega) \sim \frac{e^2}{c}\gamma_e^2\beta^2D_{\alpha\alpha}.
\label{spec_pow_diff}
\end{equation} 

Next, if we assume isotropic emission by all plasma electrons, then the jitter emission coefficient may be obtained from Eq.\ (\ref{spec_pow_diff}) with the multiplication of $n_e/4\pi$. Thus:
\begin{equation}
j^\textrm{jitter}_\omega \sim \frac{n_ee^2}{4\pi c}\gamma_e^2\beta^2D_{\alpha\alpha} = \left(\frac{m_e\omega_{pe}^2}{16\pi^2c}\right)\gamma_e^2\beta^2D_{\alpha\alpha}.
\label{jitt_emiss}
\end{equation} 

Finally, the emission coefficient for non-relativistic jitter (pseudo-cyclotron) radiation, given a Maxwellian distribution of electrons, will be: 
\begin{equation}
j^\textrm{jitter}_\omega \sim \Re[n]\left(\frac{\omega_{pe}^2k_BT_e}{2\pi^3c^3}\right)D_{\alpha\alpha},
\label{jitt_emiss_nonrel}
\end{equation} 
where we have reintroduced the index of refraction, and substituted 
\begin{equation}
\beta c=\langle |{\bf v}| \rangle = \left(\frac{8k_BT_e}{\pi m_e}\right)^{1/2}.
\label{vel_avg}
\end{equation} 
Comparing Eqs.\ (\ref{jitt_emiss_nonrel}) and (\ref{comb_emiss}), we see that they only differ by a numerical factor. Thus, Eqs.\ (\ref{jitt_emiss}) and (\ref{jitt_emiss_nonrel}) provide an attractive phenomenological definition for the ``jitter'' collision frequency, which may be obtained directly from the small-angle jitter radiation emission coefficient.

\section{Magneto-optic Effects in Small-scale Magnetic Turbulence}
\label{s:coll_eqs}

To explore the properties of electromagnetic (EM) wave propagation in quasi-collisional, magnetized plasmas, we examine the components of the dielectric tensor, $\epsilon_{ij}$. Consider an EM wave of frequency $\omega$ and with a  wave-vector ${\bf k}$, propagating through a ``cold'' magnetized plasma with an ambient magnetic field, ${\bf B}_0$. Choosing a coordinate system with $\textbf{B}_0$ parallel to the $z$-axis, and $\textbf{k}$ in the $x$-$z$ plane, the general dispersion relation is the characteristic equation \citep{sazhin93}:
\begin{equation}
\left| \begin{array}{ccc} 
-n^2\cos^2\theta + \epsilon_{xx} & \epsilon_{xy} & n^2\cos\theta\sin\theta + \epsilon_{xz}  \\ 
\epsilon_{yx} & -n^2 + \epsilon_{yy} & \epsilon_{yz} \\
n^2\cos\theta\sin\theta + \epsilon_{zx} & \epsilon_{zy} & -n^2\sin^2\theta + \epsilon_{zz}  
\end{array} \right| = 0
\label{disp_def}
\end{equation}
Where $\theta$ is the angle between $\textbf{B}_0$ and $\textbf{k}$, $n \equiv kc/\omega$ is the complex index of refraction, and
\begin{subequations}     
\begin{align}
\epsilon_{xx} = \epsilon_{yy} = {\frac{1}{2}}{(R + L)} \\ 
\epsilon_{xy} = {-}\epsilon_{yx} = {\frac{\it i}{2}}{(R - L)} \\ 
{\epsilon}_{zz} = P \\
 \epsilon_{xz} = \epsilon_{zx} = \epsilon_{yz} = \epsilon_{zy} = 0.
\end{align}
\label{comp_defs}
\end{subequations}
As a low-order approximation, collisions may be treated as drag terms, of the form $-\nu_\text{eff}{\bf v}$, in the Lorentz equation of motion for the charged plasma particles. This introduces the substitution rule: $\omega \rightarrow \omega + i\nu_\textrm{eff}$. Thus, in the cold plasma approximation, the elements of the ``collisionless'' dielectric tensor generalize to \citep{brambilla98}:
\begin{subequations}     
\begin{align}
L = 1 - \sum_{s}\frac{\omega_{ps}^2}{\omega(\omega+{\it i}\nu_s - \Omega_{cs})} \\ 
R = 1 - \sum_{s}\frac{\omega_{ps}^2}{\omega(\omega+{\it i}\nu_s + \Omega_{cs})}  \\ 
P = 1 -  \sum_{s}\frac{\omega_{ps}^2}{\omega(\omega+{\it i}\nu_s)}, \\
\end{align}
\label{comp_colls}
\end{subequations}
where $\omega_{ps}$ is the plasma frequency, $\Omega_{ps} = q_sB_0/m_sc$ is the non-relativistic gyro-frequency, $\nu_s\equiv \nu^\textrm{eff}_s$ is the effective collision frequency and the subscript $s$ denotes the plasma species (e.g., electrons and multiple ions). In our study, we will assume that only the electron dynamical time-scales are of interest; thus, $s = e$. 

The properties of EM wave propagation through a magnetized plasma depends heavily upon the orientation of the wave-vector with respect to the ambient magnetic field, ${\bf B}_0$. We will consider two limiting cases. First, we will consider propagation along the direction of ${\bf B}_0$. The difference in the indices of refraction of left- and right-circularly polarized light, as it propagate along this direction, results in the well-known {\it Faraday Effect}. As we will demonstrate, strong collisions significantly alter the conventional Faraday expressions.

\subsection{``Quasi-collisional'' Faraday Effect}

If the wave-vector is aligned with ${\bf B}_0$, the solution to Eq.\ (\ref{disp_def}) assumes the form:
\begin{equation}
\frac{c^2k^2}{\omega^2} = 1 - \frac{\omega_{pe}^2}{\omega{\sigma}\left(1 \pm \frac{\Omega_{ce}}{\sigma}\right)},
\label{fara_disp}
\end{equation} 
where $\sigma \equiv \omega + {\it i}\nu_\textrm{eff}$, and we have assumed the total collision frequency given by Eq.\ (\ref{coll_eff}). The ``$\pm$' signs refer to the right-circular and left-circular polarizations, respectively.

Next, we make the standard assumptions that $\omega \gg \Omega_{ce}$ and $\omega^3 \gg \omega_{pe}^3$. The high-order of the latter assumption is needed to keep terms (linearly) proportional to the electron number density, $n_e \propto \omega_{pe}^2$. Next, we expand Eq.\  (\ref{fara_disp}) in the small parameter, $\sigma$:
\begin{equation}
\frac{c^2k^2}{\omega^2} \approx 1 - \frac{\omega_{pe}^2}{\omega\sigma}\left[1 \mp \frac{\Omega_{ce}}{\sigma}\right].
\label{fara_disp_expan}
\end{equation} 
Expanding the square root results in the index of refraction yields:
\begin{equation}
n \approx 1 -  \frac{\omega_{pe}^2}{2\omega\sigma}\left[1 \mp \frac{\Omega_{ce}}{\sigma}\right].
\label{fara_N}
\end{equation} 

Faraday rotation is the result of the discrepancy between the wave-vectors of the two polarizations, $\Delta{k}_{\pm}$. From the real part of Eq.\ (\ref{fara_N}), we get:
\begin{equation}
\Delta{k}_{\pm} \approx \frac{\omega_{pe}^2\Omega_{ce}}{2c\left(\omega^2 + {\nu_\text{eff}}^2\right)^2.}\left[\omega^2 - {\nu_\textrm{eff}}^2\right].
\label{del_k}
\end{equation} 

The existence of an imaginary part in Eq.\ (\ref{fara_N}) indicates the presence of absorption. The absorption coefficient is given by the general relation:
\begin{equation}
\alpha_\textrm{absp} \equiv -2\frac{\omega}{c}\Im[n]
\label{absp_def}
\end{equation} 
Thus, the Faraday (quasi-)collisional absorption coefficient is: 
\begin{equation}
\alpha_\textrm{absp}^\textrm{Farad} \equiv -\frac{\omega_{pe}^2\nu_\text{eff}}{c\left(\omega^2 + {\nu_\text{eff}}^2\right)}\left[1 \mp \frac{2\Omega_{ce}\omega}{\left(\omega^2 + {\nu_\text{eff}}^2\right)}\right].
\label{absp_def2}
\end{equation} 

Finally, the total change in the polarization phase angle, $\Delta{\Psi}$ is obtained by the integration of $\Delta{k}_\pm$ along the path of the EM wave. Operationally, $\Omega_{ce}$ and $\omega_{pe}$ are functions of position, $z$. The latter depending, straightforwardly, upon the electron density, $n_e(z)$. There is subtlety in the interpretation of the gyro-frequency, however. Traditionally, it is defined here as: 
\begin{equation}
\Omega_{ce} \equiv \frac{eB_{\parallel}(z)}{m_e c},
\label{omega_ce_def}
\end{equation} 
where $B_{\parallel}(z)$ is the component of the magnetic field, at $z$, parallel to ${\bf k}$. It is implicitly assumed that ${\bf B}_0$ is super-Larmor-scale, which is an underlying assumption of the (linear) cold plasma approximation. 

Thus, the proper physical interpretation of our result is that $B_{\parallel}(z)$ refers only to the large-scale component of the magnetic field, whereas $\nu_\text{eff}$ is the result of small-scale magnetic fluctuations. Hence, using Eq. (\ref{del_k}), we may write the collision-corrected expression for the Faraday rotation angle as: 
\begin{equation}
\Delta{\Psi} = \frac{2\pi e^3}{m_e^2c^2} \int \frac{\left[\omega^2 - \nu_\text{eff}(z)^2\right]}{\left[\omega^2 + \nu_\text{eff}(z)^2\right]^2}n_e(z)B_\parallel(z) \text{d}z.
\label{fara_int}
\end{equation} 

Formally, the collision frequency may be a function of $z$; which is why we have included it in the integrand. To simplify the treatment even further, we assume a constant (or averaged) collisional frequency $\nu_\textrm{eff}^\star$ throughout the entire plasma. Then, Eq.\ (\ref{fara_int}) can be written as:
\begin{equation}
\Delta{\Psi} \simeq \frac{\left(1 - Z^2\right)}{\left(1 + Z^2\right)^2}\lambda^2 RM,
\label{fara_int_approx}
\end{equation} 
where $\lambda = 2\pi{c}/\omega$ is the radiation wavelength, $Z \equiv \nu_\textrm{eff}^\star/\omega$ is an normalized collision frequency, and 
\begin{equation}
RM \equiv \frac{e^3}{2\pi m_e^2c^2} \int n_e(z)B_\parallel(z) \textrm{d}z,
\label{RM_def}
\end{equation} 
 is the standard collisionless {\em rotation measure}. 

In the absence of (quasi-)collisions, when $Z = 1$, Eq.\ (\ref{fara_int_approx}) gives the conventional result. Thus, the ratio:
\begin{equation}
\frac{\Delta\Psi}{\lambda^2RM} = \frac{\left(1 - Z^2\right)}{\left(1 + Z^2\right)^2} =  \frac{\Delta\Psi_\text{collisional}}{\Delta\Psi_\text{collisionless}},
\label{discrep_rat}
\end{equation} 
illuminates a possible, (quasi-)collisionality-induced, discrepancy. 

\begin{figure}
\includegraphics[angle = 0, width = 1\columnwidth]{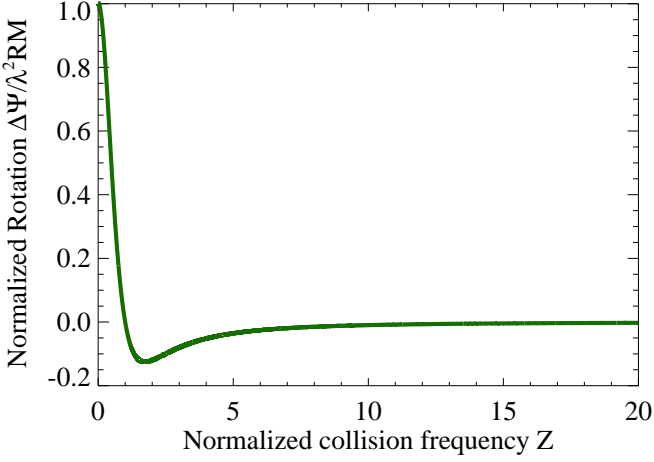}
\caption{(Color online) Normalized Faraday rotation angle vs. the normalized collision frequency. Notice that at  $Z \equiv\nu_\text{eff}^\star/\omega= 1$ zero Faraday rotation occurs. Collisions have effectively nullified Faraday Rotation.}
\label{farad_Z}
\end{figure}

In Figure \ref{farad_Z}, we have plotted Eq.\ (\ref{discrep_rat}) as a function $Z$.
The curve has a number of interesting properties. Firstly, when $Z = 1$ (i.e., $\omega = \nu_\text{eff}^\star$), zero rotation occurs. Evidently, in this case, (quasi-)collisions have effectively nullified Faraday Rotation. 

Secondly, the rotation angle remains negative for $Z > 1$; obtaining a minimum value of $-1/8$ at $Z = \sqrt{3}$. Finally, as $Z \rightarrow \infty$, the rotation angle approaches zero.

How much do standard Coulomb collisions affect Faraday rotation observations/measurements?  
For example, in the interstellar medium with density $n_e \sim 1 \ cm^{-3}$, the electron-ion collisional frequency is about $\nu_{ei} \simeq 7\times10^{-5} s^{-1}$. The strongest effect is expected at the observation frequency $\omega\sim\nu_{ei}$, which is well below any viable frequency range for Faraday polarimetry. Thus, for this reason, Coulomb collisions are generally neglected in astrophysical environments. Nevertheless, quasi-collisionality may be significant where high-amplitude electromagnetic turbulence is suspect. Thus, the observation of a Faraday rotation discrepancy (as described above) may indicate the presence of small-scale magnetic fields.

\subsection{Ordinary and Extraordinary Mode Propagation in ``Quasi-collisional'' Solid-density Laser Plasmas}

In the plane perpendicular to ${\bf B}_0$, two distinct wave modes may propagate. The first of these is the {\it Ordinary mode} (or $O$-mode), which is equivalent to the electromagnetic wave solution for a non-magnetized plasma. The index of refraction for the $O$-mode, accounting for collisions, is:   
\begin{equation}
n_O^2 = 1 - \frac{X}{1 + Z^2} + iZ\frac{X}{1+Z^2},
\label{n_O}
\end{equation} 
where $X \equiv \omega_{pe}/\omega$.
Since we cannot safely assume that $Z \ll 1$, Eq.\ (\ref{n_O}) must be solved exactly. This results in a real part \citep{ma05}:
\begin{equation}
\Re[n_O] = \frac{1}{4}{\left(\epsilon_r + \sqrt{\epsilon_r^2 + \epsilon_i^2}\right)}^2,
\label{eps_r}
\end{equation} 
and an imaginary part:
\begin{equation}
\Im[n_O] = \frac{1}{2\Re[n_O]}\epsilon_i,
\label{eps_i}
\end{equation} 
where $\sqrt{\epsilon_r} \equiv \Re[n_O]$ and $\sqrt{\epsilon_i} \equiv \Im[n_O]$. As before, the presence of an imaginary index of refraction implies absorption. Consequently, the $O$-mode absorption coefficient is given by the substitution of Eq.\ (\ref{eps_i}) into Eq.\ (\ref{absp_def}).

Notice that $\Re[n_O] > 0$, for all $\omega$. This means, physically, that the mode has no true cutoff frequency. For $Z \ll 1$, the approximate cutoff will be at the plasma frequency, $\omega_{pe}$, that is where $\Re[n_O]$ quickly approaches zero. In the general case, however, an effective cutoff may not be present.

The {\it Extraordinary mode} (or, $X$-mode) has a considerably more complicated dispersion relation. The exact solution of which is \citep{yesil08}:
\begin{equation}
\begin{aligned}
n_X^2 = 1 - \frac{X\left[\left(1-X\right)\left(1-X-Y^2\right)+Z^2\right]}{\left[1-X-Z^2-Y^2\right]^2 + Z^2\left[2-X\right]^2} \\
+ \  iZ\frac{X\left[\left(1-X\right)^2+Z^2+Y^2\right]}{\left[1-X-Z^2-Y^2\right]^2+Z^2\left[2-X\right]^2},
\label{n_X}
\end{aligned}
\end{equation} 
where $Y \equiv \Omega_{ce}/\omega$ and $X \equiv \omega_{pe}/\omega$. Due to complexity, we will not present an analytical analysis of this case.

Now, we will explore the implications of strong quasi-collisions for $O$-mode and $X$-mode propagation in laser-generated solid-density plasmas. We consider a metal target irradiated by a laser at normal incidence, with an intensity of $10^{18} \ W \ cm^{-2}$ (the threshold of relativistic intensity). Next, we estimate the relevant plasma parameters, assuming a fully ionized aluminium target ($Z_i = 13$) and a laser wavelength of $\lambda_l = 800 \ nm$. A decent estimate for the electron temperature is suggested by \citep{hatchett}:
\begin{equation}
k_B{T_e} \sim U_\textrm{pond} \sim 1 \ MeV \times \sqrt{\frac{I\lambda_l^2}{10^{19} \ [W \ cm^{-2} \ \mu{m}^2]}},
\label{scale_temp}
\end{equation} 
where $U_\textrm{pond}$ is the ponderomotive potential of the incident laser beam. Substitution of our laser parameters gives an electron temperature of $ 253 \ keV$. 

Assuming that the small-scale magnetic turbulence is the result of a Weibel-like instability, the magnetic field will roughly have the maximum value \citep{belyaev}: 
\begin{equation}
B^\textrm{Weibel}_\textrm{max} \sim \frac{m_e\omega_{pe}c}{e},
\label{weibel_max}
\end{equation} 
which is consistent with the theoretical saturation condition $\Omega_{ce} \sim \omega_{pe}$.

Next, we must select a model for the plasma frequency profile. We suppose an exponential profile for the electron density in the direction of the laser beam, i.e.,
\begin{equation}
n_e(z) = n_ce^{(z/\lambda_l - 1)},
\label{n_profile}
\end{equation} 
where $n_c \equiv m_e\omega^2/4\pi e^2$ is the {\em collisionless} critical electron density, and $z$ is along ${\bf k}$. We furthermore assume that the density is uniform in the transverse plane. From this profile, we choose $\omega_{pe}(z = 0)$ for substitution into Eq.\  (\ref{weibel_max}). The result is a magnetic field, $B^\textrm{Weibel}_\textrm{max} \approx 81.2 \ MG$. We will suppose the existence of a large-scale magnetic field in the metal target. For simplicity, we assume that this field is approximately uniform, and that it is situated perpendicular to the angle of normal incidence, which is typical of the laser-induced (ordered) Biermann battery fields seen in ICF experiments, although these fields assume a more complex azimuthal profile \citep{huntington15}.

Additionally, we suppose that $B_0 = B^\textrm{Weibel}_\textrm{max}$, and treat $\delta{B}$ (the small-scale component) as a free parameter.

Furthermore, the electron-ion collisions are computed using Eq.\ (\ref{nu_ei_def}), that is we ignore any non-uniformity in the electron temperature.

Lastly, we consider an effective pitch-angle diffusion coefficient for the entirety of the target. We assume that $\lambda_B \sim \lambda_l$, since for Weibel magnetic fields: $\lambda_B \sim d_e$, where $d_e = c/\omega$ is electron skin-depth at the critical surface. In practice, the correlation length should be significantly shorter than the laser wavelength, so that Eq.\ (\ref{size_def}) will hold.

\begin{figure}
\includegraphics[angle = 0, width = 1\columnwidth]{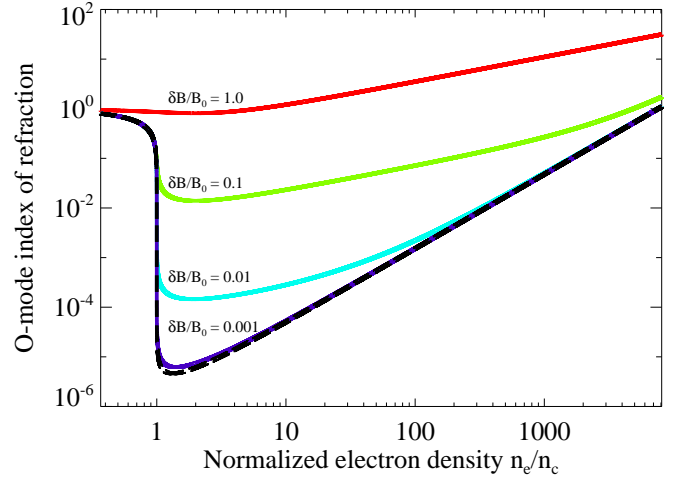}
\caption{(Color online). Index of refraction for the $O$-mode as a function of depth (in terms of the electron density). Displayed here are five solutions, all differing by the ratio, $\delta{B}/B_0$. Notice that for $\delta{B}/B_0 = 0.001$, $\nu_{ei} \gg D_{\alpha\alpha}$, and the expected weakly-collisional dependence is realized; i.e., a steep approach of the index of refraction towards zero at $n_c$. In contrast, $\delta{B} = B_0$ leads to a virtually transparent target. Included in this plot is the solution for $\nu_\text{eff} = \nu_{ei}$  --- the dashed black line.}
\label{index_dBO}
\end{figure}

In Figure \ref{index_dBO}, five solutions for the $O$-mode index of refraction are plotted as a function of the depth into the target (represented by the electron density). These solutions differ by the assumed $\delta{B}$. The effective quasi-collision frequency is significantly large for $\delta{B} \sim B_0$: $\nu_\text{eff} \approx  3.3\times10^{15} \ s^{-1}$, which is comparable to the laser frequency. This is in stark contrast to the much weaker electron-ion contribution: $\nu_{ei} \approx 7.1\times10^9 \ s^{-1}$, at the critical surface, $n_c$.  

For $\delta{B}/B_0 = 0.001$, $\nu_{ei} \gg D_{\alpha\alpha}$, and the expected weakly-collisional dependence is realized. Here, there is a steep drop in the index of refraction towards zero near $n_c$. Physically, this indicates that most of the $O$-mode wave is reflected back from the critical surface -- as, otherwise, anticipated. 
As the effective collision frequency increases, the reflectivity at the critical surface quickly drops. In fact, when $\delta{B}/B_0 = 1$, the entirety of the metal target is virtually transparent.  

The steep increase in the index of refraction, for all the curves, at high-density is a result of the density dependence in Eq. \ (\ref{nu_ei_def}). Since the metal target is of limited extent, this asymptote of the solution may not be experimentally viable. 

Next, the $X$-mode has a considerably more complicated index of refraction. The {\em collisionless} dispersion relation includes two cutoff frequencies and a resonance. 
\begin{figure}
\includegraphics[angle = 0, width = 1\columnwidth]{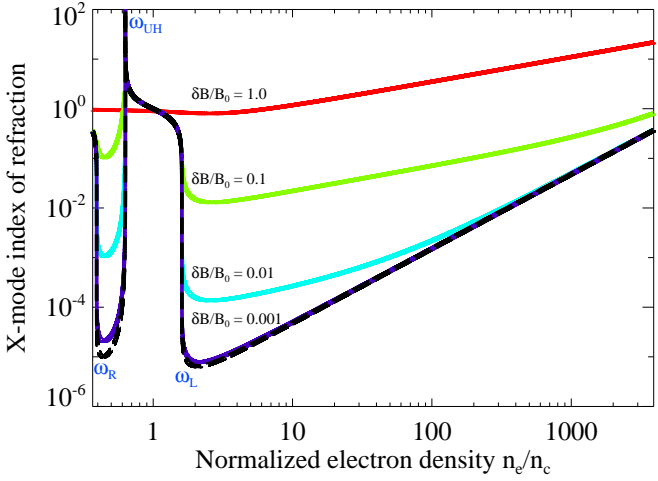}
\caption{(Color online). Index of refraction for the $X$-mode as a function of depth (in terms of the electron density). Displayed here are five solutions, all differing by the ratio, $\delta{B}/B_0$. Notice that for $\delta{B}/B_0 = 0.001$, $\nu_{ei} \gg D_{\alpha\alpha}$, and the expected weakly-collisional dependence is realized; i.e., a steep approach of the index of refraction towards zero at $\omega_\text{R}$, a resonance at $\omega_\text{UH}$, and another cutoff at $\omega_\text{L}$. Collisions effectively connect the cutoff frequencies to the resonance, allowing access to $\omega_\text{UH}$ and $\omega_\text{L}$. Nonetheless, for $\delta{B} \sim B_0$, the cutoffs and resonance disappear completely. Included in this plot is the solution for $\nu_\text{eff} = \nu_{ei}$  --- the dashed black line.}
\label{index_dBX}
\end{figure}
The first cutoff, 
\begin{equation}
\omega_\text{R} = \frac{1}{2}\left(\Omega_{ce} + \sqrt{\Omega_{ce}^2 + 4\omega_{pe}^2}\right),
\label{omega_R}
\end{equation} 
is slightly less than $\omega$. Its presence, as the first steep drop in the index of refraction, can be seen Figure \ref{index_dBX}. Next, a resonance occurs at the {\em upper-hybrid} frequency, i.e.
\begin{equation}
\omega_\text{UH} = \sqrt{\omega_{pe}^2 + \Omega_{ce}^2}.
\label{omega_UH}
\end{equation} 
The upper-hybrid resonance, similarly, occurs slightly prior to $n_c$ (see Figure \ref{index_dBX}). Lastly, a second cutoff frequency occurs at:
\begin{equation}
\omega_\text{L} = \frac{1}{2}\left(-\Omega_{ce} + \sqrt{\Omega_{ce}^2 + 4\omega_{pe}^2}\right),
\label{omega_L}
\end{equation} 
which is slightly beyond the critical surface, $n_c$.

The behavior similar to the $O$-mode profile may be observed in Figure \ref{index_dBX}. What is noteworthy here is that collisions essentially connect the cutoff frequencies to the resonance, allowing access by $\omega_\text{UH}$ and $\omega_\text{L}$. Nonetheless, when quasi-collisions dominate the dispersion, as they do for $\delta{B} \sim B_0$, the cutoffs and resonance disappear completely.

\begin{figure}
\includegraphics[angle = 0, width = 1\columnwidth]{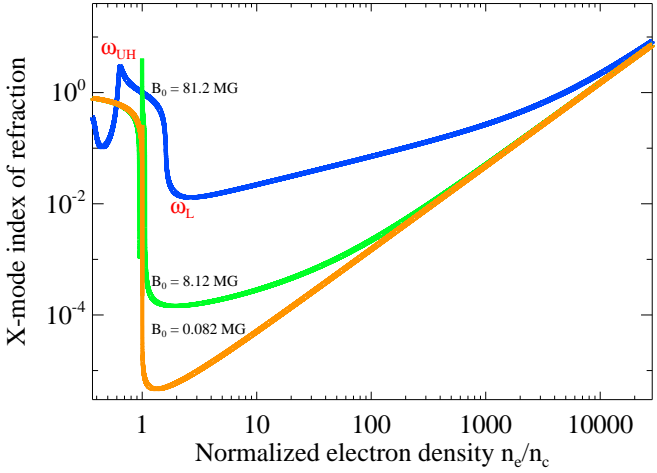}
\caption{(Color online) Index of refraction for the $X$-mode as a function of depth (in terms of the electron density). Here, three solutions for which $\delta{B}/B_0 = 0.1$ are plotted with a variable $B_0$. As expected, the solution approaches the $O$-mode profile for $B_0 \rightarrow 0$.}
\label{index_B}
\end{figure}

Next, the $X$-mode index of refraction depends upon the ambient magnetic field via $\Omega_{ce}$. In Figure \ref{index_B}, we have plotted three solutions for which $\delta{B}/B_0 = 0.1$, but $B_0$ differs by orders of magnitude.
As expected, the solution approaches the $O$-mode profile for $B_0 \rightarrow 0$.

Finally, the quasi-collisional absorption is a very important consideration as well. Ignoring reflection and refraction, the intensity, $I$, falls off exponentially while traversing a lossy medium, i.e.,
\begin{equation}
I(z) = I_0e^{-\int |\alpha_\text{absp}(z)| \text{d}z},
\label{beer_law}
\end{equation} 
where $I_0$ is the vacuum intensity.
\begin{figure}
\includegraphics[angle = 0, width = 1\columnwidth]{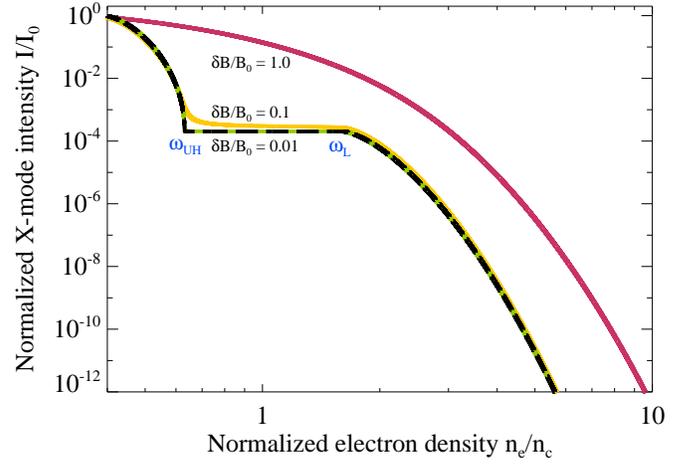}
\caption{(Color online). $X$-mode laser intensity as a function of the target depth (in terms of the electron density). Despite the relative transparency for $\delta{B} \sim B_0$, the laser intensity quickly decays beyond the critical surface. Interestingly, the laser intensity is relatively fixed from $\omega_\textrm{UH}$ to $\omega_{L}$, for low-$Z$. Additionally, there is initial drop near $\omega_\textrm{UH}$ that is not present in the high-$Z$ case. Included in this plot is the solution for $\nu_\text{eff} = \nu_{ei}$  --- the dashed black line.}
\label{absorp_dBX}
\end{figure}
In Figure \ref{absorp_dBX}, we have used Eqs.\ (\ref{absp_def}) and (\ref{beer_law}) to plot the $X$-mode intensity as a function of depth for the same conditions as in Figure \ref{index_dBX} (excluding $\delta{B}/B_0 = 0.001$).

Despite the relative transparency of the plasma for $\delta{B} \sim B_0$, Figure \ref{absorp_dBX} shows that the laser intensity quickly decays beyond the critical surface. Interestingly, the laser intensity is relatively fixed from $\omega_\text{UH}$ to $\omega_{L}$, for low quasi-collisionality, i.e., low-$Z$. 
\begin{figure}
\includegraphics[angle = 0, width = 1\columnwidth]{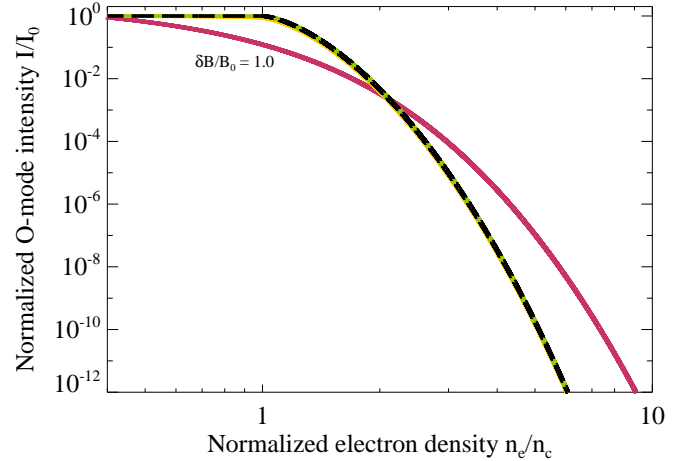}
\caption{(Color online). $O$-mode laser intensity as a function of the target depth (in terms of the electron density). Once more, we see a relatively fixed laser intensity up to the critical surface, for low-$Z$. The high-$Z$ curve is identical to the $X$-mode equivalent. Included in this plot is the solution for $\nu_\text{eff} = \nu_{ei}$  --- the dashed black line.}
\label{absorp_dBO}
\end{figure}
Figure \ref{absorp_dBO} displays the same scenario for the $O$-mode case. Once more, we see a relatively fixed laser intensity up to the critical surface, for low-$Z$. The high-$Z$ (i.e., $\delta{B} \sim B_0$) curve is identical to the $X$-mode equivalent, thus demonstrating the dominance of quasi-collisions over the ``magnetization'' effect from $B_0$ at large $\delta{B}$.

From Figures \ref{index_dBO}-\ref{absorp_dBO}, it is clear that effective quasi-collisionality in solid-density laser plasmas may be significant. Although the high-$Z$ scenario of $\delta{B} \sim B_0$ is unlikely, the presence of small-scale magnetic fields (especially near the critical surface) may, unanticipatedly, impact the reflectivity and absorption. The effect may be critically important to certain setups, such as the inertial confinement fusion (ICF) experiments or experiments that exploit the Cotton-Mouton effect for magnetic field diagnostics.

\section{Conclusions}
\label{s:concl}

In this paper, we have investigated the implications the quasi-collisionality induced by small-scale magnetic turbulence in, otherwise, collisionless plasma environments. Our results demonstrate that radiative transport is dramatically affected by the presence of strong effective collisions.

Particularly, our analysis shows that sub-Larmor-scale magnetic fluctuations in magnetized plasmas may sharply attenuate Faraday rotation measures ($RM$). In fact, with the effective quasi-collision frequency on the same order as the wave frequency, the Faraday rotation effect may be completely canceled, hence $RM = 0$. In an unexpected turn, with $\nu_\text{eff} > \omega$, we predict {\em negative} $RM$ values in these environments. These results are crucial for Faraday rotation-based laboratory plasma diagnostics and interpretation of the results of astronomical observations of Faraday rotation measures of magnetized astrophysical and space plasmas, e.g., of the interstellar and intracluster media.

In the laboratory setting, we find that small-scale turbulence may complicate the propagation of EM waves through high-intensity laser-plasmas; specifically, solid-density laser-plasmas. Namely, the reflectivity and absorption of $X$- and $O$-modes is largely affected when the plasma is highly ``collisional''. In fact, for sufficiently high (quasi-)collisionality, the plasma cutoff frequencies cease to exist. 

These effects can have crucial implications for the ICF performance. Indeed, the high quasi-collisionality regime occurs when the Weibel instability or other kinetic filamentation instabilities are excited to produce strong sub-Larmor magnetic (or possibly fully electro-magnetic) fields. In this regime, the plasma may happen to be transparent so that the critical surface ceases to exist. The impulse delivered to the imploding plasma by radiation pressure halves in the case (cf. reflection vs. absorption), which greatly affects ICF performance. For the same reason, the absorption coefficient reduces too, so that the depth through which radiation can penetrate into the target increases, which changes the energy deposition profile in the target. How theis affect the ICF performance remains to be seen from dedicated theoretical analyses and numerical simulations. On the other hand, we stress that the performance, being affected by quasi-collision-induced transparency which depends on $\delta B/B_0$, can be controlled by the ambient magnetic field, $B_0$, both via the Weibel instability suppression (by lowering $\delta B$) and the reduction of the effective quasi-collisionality of the plasma (by increasing $B_0$ for a fixed $\delta B$). 

We should also mention that the role of small-scale electric fields (of the order of the skin depth, as in Langmuir turbulence, for example) has not been investigated here. However, the scattering effect of such fields is expected to be similar to the magnetic fields, although the particle energy may no longer be constant in scatterings. Thus, we expect the electrostatic and fully electromagnetic fields to result in qualitatively similar effects, though quantitative predictions may differ.

We propose that quasi-collisional magneto-optic effects may be exploited for diagnostic purposes. Since the effective quasi-collision frequency --- the pitch-angle diffusion coefficient, Eq. \ (\ref{Daa}) --- is proportional to the magnetic field correlation length and the square of the small-scale magnetic fluctuations, it provides a novel means by which the statistical properties of the small-scale magnetic turbulence may be identified. Additionally, the jitter radiation spectrum readily provides a phenomenological definition for the effective collision frequency, \emph{\`{a} la} Eq.\ (\ref{jitt_emiss}). Jitter radiation may be directly observable in several of these plasma environments, e.g., high-intensity solid-density laser plasmas \citep{keenan15b}.

Our model, nonetheless, has some limitations. In particular, strong sub-Larmor-scale magnetic fluctuations are not likely present in all collisionless or weakly collisional plasmas. Leading candidates for the existence of strong fluctuations include: collisionless shocks in gamma-ray bursts and early moments of supernova explosions, high-intensity laser plasmas, and turbulent solar wind and magnetosphere/magnetotail plasmas. Our principal assumption that the system spatial scale is much greater than the small-scale magnetic correlation length seems to rule out most interstellar and intergalactic plasmas, where the magnetic correlation lengths are believed to be $\sim 100 \ pc$ and $\sim kpc-Mpc$, respectively \citep{beck87, neronov13}. Allowing for hidden small-scale components (with smaller correlation lengths) in these environments requires unrealistically large magnetic fields to keep the absorption {\it e-folding} distance at parsec to kiloparsec scales; this is required so that a signal may not be completely absorbed in transit.

To conclude, the obtained results suggest that small-scale magnetic fluctuations conceal a “collisional” signature, which may provide a useful radiative diagnostic of magnetic micro-turbulence in laboratory, astrophysical, space and solar plasmas, as well as significantly affect performance of inertial confinement fusion and laser plasma experiments.

\begin{acknowledgments}
This work was partially supported by the DOE grant DE-FG02-07ER54940 and the NSF grant AST-1209665.
\end{acknowledgments}

\end{document}